\documentclass{aa}

\usepackage{graphicx}

\usepackage{txfonts}

\usepackage{natbib}

\begin{document}
 
\title{The parsec-scale properties of the radio galaxy 4C 26.42 in the dense cooling core cluster A1795.}

\author{E. Liuzzo
          \inst{1,2}
          \and G. B. Taylor\inst{3}
          \and G. Giovannini\inst{1,2}
          \and M. Giroletti\inst{1}
          }

   \offprints{liuzzo@ira.inaf.it}

   \institute{Istituto di Radioastronomia, INAF, via Gobetti 101, 40129 Bologna, Italy.\\
         \and
            Dipartimento di Astronomia, Universit\`a di Bologna, via Ranzani 1 , 40127 Bologna, Italy\\
         \and
            Department of Physics and Astronomy, University of New Mexico, Albuquerque NM 87131, USA; Greg Taylor is also an Adjunct Astronomer at the National Radio Astronomy Observatory\\ 
             }

   \date{Received March 3, 2009; accepted May 5, 2009}

\abstract{}{The aim of the present work is to study the radio
   emission on the parsec scale of 4C 26.42 , the Brightest cluster galaxy in Abell 1795, in the 
   framework of radiosources in a dense cool core cluster.}
{We present Very Long Baseline Array (VLBA) observations at
  1.6, 5, 8.4 and 22 GHz. We performed a spectral index and multiepoch analysis. }
{ The source appears two-sided with a well defined and
  symmetric Z-structure at $\sim$5 mas from the core. The
  kiloparsec-scale morphology is similar to the parsec-scale
  structure, but reversed in P.A., with symmetric 90$^\circ$ bends at about 2 arcsec
  from the nuclear region.  Comparing data obtained at 3 different
  epochs we derive a 3$\sigma$ limit to the apparent proper motion of
  $\beta_a$ $<$ 0.04.  We suggest that the parsec-scale jets are
  sub-relativistic, in contrast with the high velocities found for most
  low-power radio galaxies.  The origin of the unusual radio
  morphology remains a puzzle. We suggest that the identification of
  the parent galaxy with the central cD in a cooling cluster plays an
  important role in the properties and structure of the jets}
{}


\maketitle

\section{Introduction}

Studies of central cluster galaxies show that amorphous and/or
distorted radio sources are typical of the cooling cluster environment \citep{ode87,tay94b}.  
Recently, using the high resolution imaging
capabilities of the X-ray satellite Chandra, the interaction between
the hot gas and the radio plasma in cooling clusters has become
clearer, and it is now accepted that the intracluster gas and the
central radio galaxy profoundly influence each other.  Examples
include Hydra~A \citep{mcn00} 
and Perseus~A \citep{fa00}. 
In both cases, there is an anti-coincidence between the radio lobes and X-ray
emission. In particular, the kiloparsec radio lobes are located in
X-ray cavities. Moreover \citet{bau91} 
demonstrated that the central
radio sources can arrest or slow down the cooling process. The central
AGN can inject into the intra-cluster medium (ICM) an amount of energy
comparable to that lost by the cooling X-ray gas.

High rotation Faraday measures (RMs) are frequently found toward radio
sources at the centers of cooling core clusters.  This is a direct
consequence of the thermal gas and cluster magnetic fields present in
the ICM. The detection of molecular gas (e.g., CO) and H$_{\alpha}$
emission confirms the presence in the ICM of dense cold gas as well.
\citet{bau91} and \citet{sa95} discussed a possible interaction between the radio
sources and the cooling flow. For example, the disruption of the radio
jets can be the consequence of the high-pressure ambient gas, which
also explains the plume-like features of some sources.

At parsec-scale, there are well known sources in cooling core clusters
like Hydra~A (Taylor, 1996), 3C84 (Taylor et al., 2006), PKS 1246-410 in Centaurus cluster (Taylor et al., 2006), and PKS 2322-123 in A2597 \citep{tay99}. All but one of these (PKS 1246-410) show two-sided
structures and mildly relativistic jets. This is in sharp contrast to
the predominately one-sided parsec-scale jets found in most 
low power radiogalaxies \citep{gio01}.

The radio galaxy 4C 26.42 (B2 1346+26) is identified with a very
luminous (M$_{v}$=-23) cD galaxy at z=0.0633. This cD galaxy is the
brightest member of the rich cluster Abell 1795.  In X-rays, A1795
shows a relaxed structure and it is one of the most powerful clusters
of galaxies known, with a luminosity of $\sim$8.5 $\times$ 10$^{44}$
ergs s$^{-1}$ in the 0.5-4.5 keV band. A more recent Chandra study of \citet{et02} indicate that the central 200-kpc region is undergoing significant cooling with a gas mass deposition rate of about 100 M$_{\odot}$ yr $^{-1}$ in the absence of any heating process. They also found that the shape of the potential is in agreement with the motion of the central dominant galaxy, and suggests that the central cluster region is not relaxed. 
It has a strong central peak of
cooler gas and this evidence of ``cooling flow'' \citep{bau92}
 is supported by the presence of strong emission-line nebulosity \citep{ca00, fa94} 
around the central cD galaxy along with excess
blue light. The blue light is probably due to massive stars \citep{mcn96} 
with some contribution from young globular stars \citep{holt96}.
 Molecular gas in the cD galaxy has been detected through
molecular hydrogen emission \citep{fak98}.
\citet{fa94} observed in the core an X-ray filament that coincides with an H$_{\alpha}$
emission feature discovered by \citet{co83}. 
This filament is produced by cooling gas from the ICM, and could be significant
in improving our understanding of the energy and ionization source
of the optical nebulosity, so frequently found in cooling core clusters.

High Faraday rotation measures (exceeding $\sim$2000 rad m$^{-2}$ in
places) are found in 4C 26.42 in high resolution (0.6 arcsec) VLA images \citep{ge93}. 
The magnitude and scale of the RM are consistent with a hot ($\sim$10$^{8}$) and dense
($\sim$0.03 cm$^{-3}$) X-ray emitting gas. The strength of the
magnetic field is probably between 20 and 100 $\mu$G, depending on the
degree of ordering.

From Very Large Array (VLA) observations, 4C26.42 is known to have an
FR I, Z-shaped small (2$''\times$12$``$) radio morphology \citep{van84}, with a total power at 0.4 GHz Log$P_\mathrm{tot,\, 0.4\, GHz}$=25.47 W/Hz and a core power at 5
GHz Log$P_\mathrm{core,\,5\, GHz}$= 23.70 W/Hz \citep{gio05}.

On the parsec-scale, \citet{gio05} observed 4C 26.42 at 5 GHz with the
Very Long Baseline Array (VLBA).  They found a faint peculiar
structure with a total flux density of $\sim$8 mJy, but the observations were inconclusive because of its low
brightness.  In this paper we discuss new multi-epochs, multi-frequencies  
VLBA observations of this source using the phase-referencing mode. 
In \S2 we present the new data, in \S3 we discuss the nuclear
properties of this source. In \S4 we discuss possible origins for
the peculiar radio structures.

Throughout this paper, we assume H$_{0}$= 70 km s$^{-1}$ Mpc$^{-1}$, 
$\Omega_{m}$ =  0.3, $\Omega_{\Lambda}$ = 0.7. At the distance of 4C 26.42, 
1 mas corresponds to 1.2 pc.

\section{Observations and data reduction.}

We obtained new VLBA observations at 1.6 GHz on 2003 August 07, at 5
GHz on 2005 July 27, and at 1.6, 5, 8.4 and 22 GHz on 2008 February 26
(see Tab.\ref{tab:cal}).  Moreover we re-reduced the data at 5 GHz of 1997 April
4 \citep{gio05}.  
All data were obtained using the phase referencing
mode, except for the 5 GHz data of 1997, whose quality largely
improved after applying the improved source position obtained from the
newer data taken using phase referencing. High resolution and high frequency data are necessary to study the nuclear region. Data at 1.6 and 5 GHz have the sensitivity and (u-v) coverage necessary to study the more extended jet structure.

\begin{table}
\caption{Calibrators list.}
\label{tab:cal}
\centering
\begin{tabular}{cccl}
\hline\hline
Epoch & Frequency & Observing time & phase calibrators\\
yy-mm-dd & GHz & hour&\\
\hline\hline
03-08-07&1.6&1.0&J1350+3034\\
08-02-26&1.6&1.0&J1342+2709 \\
97-04-06&5.0&1.0&J1350+3034\\
05-07-27&5.0&1.5&J1342+2709\\
08-02-26&5.0&1.5&J1342+2709\\
08-02-26&8.4&2.0&J1342+2709\\
08-02-26&22&3.5&J1342+2709\\
\hline
\end{tabular}
\end{table}

The observations were correlated in Socorro, NM. A total of 16 MHz
bandwidth was recorded in both right and left circular
polarization. Post-correlation processing used the NRAO AIPS package
and the Caltech Difmap package. We followed the same scheme for the
data reduction of all data sets. Using AIPS we applied ionospheric
corrections and corrections to the Earth Orientation Parameters
(EOPs). After this, we used the AIPS script VLBACALA to correct
sampler offsets and to apply a-priori amplitude calibration.  With
VLBAPANG we corrected the antenna parallactic angles and with VLBAMPCL
we removed the instrumental delay residuals. All calibrator data were
also globally fringe-fitted and solutions were applied to the target
sources with VLBAFRGP. After flagging bad visibilities, we obtained
good models for the calibrators, which we used to improve the
amplitude and phase calibration for the entire data set. Final maps
were obtained with Difmap after averaging over IFs and in time. After
editing, we applied multiple iterations of imaging and
self-calibration in phase and amplitude. At last, to these improved quality data, we applied Modelfit to obtain the final set of components describing the data.

In Tab. \ref{tab:par} we report the image parameters for final natural weighting maps at different epochs and frequencies.

\begin{table*}
\caption{Image Parameters for natural weighting maps.}
\label{tab:par}
\centering
\begin{tabular}{ccccccc}
\hline\hline
Epoch &Frequency & Beam Size & Beam P.A. & Noise &
Peak & Total Flux Density\\
yy-mm-dd & GHz & mas & deg & mJy beam$^{-1}$ &
mJy beam$^{-1}$ & mJy\\
\hline\hline
03-08-07&1.6&11.0$\times$6.1&7&0.15&22.5&67.6\\
08-02-26&1.6& 11.3$\times$5.9&15&0.10&32.2&90.9\\
97-04-06&5&4.0$\times$2.4&6&0.21&8.7&29.9\\
05-07-27&5&3.4$\times$2.2&-8&0.13&10.0&42.0\\
08-02-26&5&3.0$\times$1.5&-2&0.16&7.9&44.1\\
08-02-26&8.4&1.9$\times$0.9&-4&0.08&7.0&25.2\\
08-02-26&22&1.0$\times$0.7&-19&0.15&6.6&13.6\\
\hline
\end{tabular}
\end{table*}

\section{Results}

\subsection{The Parsec Scale Structure.}

In Figs.1-3 we present the full resolution images of 4C 26.42 at 1.6,
5, 8 and 22 GHz respectively.  At low frequency and low resolution,
the source shows two components (A and B) in the inner 15 mas (in PA
$\sim$ 60$^\circ$) and two symmetric lobes oriented North - South (N1
and S1) with a peculiar Z shaped structure (Fig.~1, left).  The source size is $\sim$ 30 mas.  At 5
GHz the two central components of the low resolution images are
resolved into three components (A, C and B; see Fig.~1, right). The symmetric
S1 and N1 lobes are still visible though with an extension of only
$\sim$10 mas.  We note that components A and B are coincident with the
regions where the jet position angle changes by $\sim$90$^\circ$.  The
parsec-scale structure is very symmetric.  At still higher resolution
(Fig.~2), we distinguish four components in the central 10 mas (A1,
A2, C and B); N1 is completely resolved out while S1 is marginally detected.  At the highest resolution, only the central
components A1, A2 and C are visible (Fig.~3). We note that in high resolution images (8.4 and 22 GHz) because of the low signal to noise ratio the extended source structure is not visible in our images.

Comparing the parsec and kiloparsec structure (Fig.~4), we note that
two changes in P.A. occur, one at $\sim$15 mas from the core where
P.A. goes from 60$^{\circ}$ to $-$30$^{\circ}$ and the
se\-cond at $\sim$ 2 arcsec where P.A. goes from $-$30$^{\circ}$ back to
60$^{\circ}$ to realign with the inner parsec-scale P.A.

\subsection{Modelfit results.} \label{}

Modelfitting was applied to all the data available for this source
using elliptical, circular or delta components. The difmap  mo\-delfit program fits aggregates of various forms of model components, fitting directly to the real and imaginary parts of the
observed visibilities using the powerful Levenberg-Marquardt
non-linear least squares minimization technique. First we applied
modelfitting to the last observations taken in 2008 independently for
each frequency. All parameters of modelfit components were initially allowed to
vary freely, however to facilitate comparisons between epochs,
a mean shape was fixed for each frequency and only the position
and flux density of the component was allowed to vary. We have also tried to describe elongated features with several small components. However, this resulted in an increase of $\chi^{2}$, so we retained the simplest model. For example, in the case of S1 at 5 GHz (1997 April 6), the $\chi^{2}$ increases from 1.393 (one component) to 1.709 (3 components). Following this step, we supposed that only the flux density of the components can change with the time. We fixed the shape of the components and then modelled the ($u$,$v$) data from previous epochs at the
corresponding frequencies. 

In Tab. \ref{tab: modelfit}, we summarize the results obtained. For high frequencies
($>$ 1.6 GHz) we refer the component distance to component C because
of symmetry reasons and its identification as the `core' of 4C 26.42
(see Sect. 3.4). At 1.6 GHz where the central component C is not
visible, we use component A as the reference point.  In Table 3 we give
the frequency (col.1) and the epoch of observation (col.2), the $\chi^{2}$ (col. 3) of the fit, the name of the components (col. 4),
polar coordinates (r) and ($\theta$) (col. 5 and col. 6) of the center
of the component relative to the origin discussed above, with polar
angle measured from the north through east, the positional uncertainty $\Delta$ (x,y) (col.7), the major axis $a$
(col. 8), the ratio $b/a$ of minor $b$ and major $a$ axes (col. 9)
of the FWHM contour, the position angle $\Phi$ (col. 10) of the major
axis measured from north to east and the flux density $S$ (col. 11). Statistical errors are provided for the component parameters in Table 3. For elliptical and circular gaussian componentes we also calculated errors in size ($a$). Uncertainties in the sizes, positions and fluxes for components were derived from signal-to-noise ratios and component
sizes \citep{form99}.  True errors could be larger in the event that
some components are covariant.

\begin{table*} [t!]
\caption{Results of Modelfit applied to 1.6, 5, 8.4 and 22 GHz data.}
\centering
\label{tab: modelfit}
\centering
\begin{tabular}{c c c c c c c c c c c}
\hline\hline
Frequency&Epoch &$\chi^{2}$&Component      & r &$\theta$  &$\Delta$ (x, y)&a  &b/a & $\Phi$ & S \\
GHz&yy-mm-dd&  && mas &deg  & mas&mas  &  & deg & mJy\\
\hline\hline
&&&&&&&&&&\\
1.6     &03-08-07    &1.312   &A, e   &0  &0 &$-$  & 3.29$\pm$0.02& 0.5&29.6&24.9$\pm$1.3\\
&&&N1, e&7.61&-130&0.02& 6.20$\pm$0.04& 0.5&-67.3&21.4$\pm$1.1\\
&&&B, e&10.58&-22&0.05&8.88$\pm$0.10&0.6&-14.7&13.1$\pm$0.7\\
&&&S1, c&14.11&-165&0.35&4.23$\pm$0.07 &1&-84.2&8.2$\pm$0.4\\
1.6&08-02-26&1.001&A, e&0&0&$-$ &3.29$\pm$0.01& 0.5&29.6&31.9$\pm$1.6\\
&&&N1, e&8.08&-128&0.01 &6.20$\pm$0.02& 0.5&-67.3&31.0$\pm$1.6\\
&&&B, e&10.18& -17&0.02 &8.89$\pm$0.04&0.6&-14.7&21.4$\pm$1.1\\
&&&S1, c&16.01&-178&0.03 & 4.23$\pm$0.06 &1&-84.2&6.6$\pm$0.3\\
&&&&&&&&&&\\
\hline
&&&&&&&&&&\\
5&97-04-06&1.393&C, e&0&0&$-$&1.92$\pm$0.05&0.3&45.7&7.6$\pm$0.4\\
&&&A, e&4.02&64&0.15&1.43$\pm$0.03&0.9&44.7&10.4$\pm$0.6\\
&&&N1, e&11.19&-6&0.55&10.38$\pm$1.09&0.4&-40.5 &2.0$\pm$0.2\\
&&&B, e&5.19&-130&0.03&2.21$\pm$0.06&0.7&30.9&7.3$\pm$0.4\\
&&&S1, e&9.4&-174&0.43&11.08$\pm$0.86&0.3&-30.6&2.7$\pm$0.2\\
5&05-07-27&1.205&C, e&0&0&$-$&1.92$\pm$0.03&0.3&45.7&8.3$\pm$0.4\\
&&&A, e&4.02&65&0.01&1.43$\pm$0.02&0.9&44.7&11.9$\pm$0.6\\
&&&N1, e&12.04&-4&0.13&10.38$\pm$0.25&0.4&-40.5&5.3$\pm$0.3\\
&&&B, e&5.07&-129&0.02&2.21$\pm$0.03&0.7&31.0&8.4$\pm$0.4\\
&&&S1, e&7.38&-158&0.09&11.08$\pm$0.18&0.3&-30.6&8.1$\pm$0.4\\
5&08-02-26&0.946&C, e&0&0&$-$&1.92$\pm$0.03&0.3&45.7&9.3$\pm$0.5\\
&&&A , e&4.02&64&0.01&1.43$\pm$0.02&0.9&44.7&11.5$\pm$0.6\\
&&&N1, e&11.10&-4&0.13&10.38$\pm$0.23&0.4&-40.5&5.9$\pm$0.3\\
&&&B, e&5.04&-127&0.02&2.21$\pm$0.04&0.7&30.9&7.7$\pm$0.4\\
&&& S1, e&6.85&-150&0.08&11.08$\pm$0.15&0.3&-30.6&9.7$\pm$0.5\\
&&&&&&&&&&\\
\hline
&&&&&&&&&&\\
8.4&08-02-26&0.890&C, c&0& 0&$-$&0.34$\pm$0.01&1&-11.9&5.3$\pm$0.3\\
&&&A1, d&3.01&61&0.02&0&1&0&2.9$\pm$0.2\\
&&&A2, c&4.02& 63&0.01&0.44$\pm$0.01&1&-14.7&8.1$\pm$0.4\\
&&&B, e&5.09&-129&0.01&2.21$\pm$0.02&0.6&11.0&7.5$\pm$0.4\\
&&&S1, c&8.89 &177&0.24&11.08$\pm$0.47&0.3&-25.6&1.9$\pm$0.1\\
&&&&&&&&&&\\
\hline
&&&&&&&&&&\\
22&08-02-26&1.040&C, c & 0 &0&$-$&0.9$\pm$0.03&1&-25.0&4.6$\pm$0.3\\
&&&A1, d& 2.05 & 71&0.01&0&1&0&2.2$\pm$0.2\\
&&&A2, c&4.04&65&0.01&0.15$\pm$0.01&1&-21.8&6.8$\pm$0.4\\
&&&&&&&&&&\\
\hline\hline
\multicolumn{11}{c}{\scriptsize Parameters of each component of the model brightness 
distribution: }\\
\multicolumn{11}{l}{\scriptsize col.1: frequency;  col.2: epoch; col.3: ($\chi^{2}$) of the fit; col.4 : name of each component and we use (e) for  elliptical gaussian component, (c) is for circular
}\\ 
\multicolumn{11}{l}{\scriptsize  gaussian component, (d) for delta component; col. 5: (r) and ($\theta$), polar coordinates of the center of the component relative to an arbitrary origin, with
 }\\
\multicolumn{11}{l}{\scriptsize  polar angle measured from the north through east; col.6: $\Delta$ (x,y) position uncertainty, ($-$) indicates no error can be determined; col.7: (a) major axis; 
}\\
\multicolumn{11}{l}{\scriptsize col.8: (b/a) ratio between minor (b) and major (a)  axes of the FWHM contour; col.9: ($\Phi$), position angle of the major axis measured from north to east;
}\\
\multicolumn{11}{l}{\scriptsize col. 9: (S) flux density.
}\\
\end{tabular}
\end{table*}

We note that the total flux density at 5 GHz, obtained by summing the
different components in the 2008 epoch, is 44 mJy, therefore
neglecting variability, only a small fraction ($\sim$ 17$\%$) of the
flux density is lost with respect to the sub-arcsecond core flux
density (53 mJy) obtained from VLA data by \citet{van84}. Lower total flux density at 5 GHz of data carried out on 6 April 1997 are due to different not in phase referencing mode (see \S2) observational analysis and it is not consequence of variability in the radioemission of the source. 
We compared the flux density of components A, B and C at different epochs at 5 GHz
to search for possible variability (see Table 3). The flux density is
slightly different but no clear trend is present. The limit on the
amount of variability is $\sim$0.6 mJy at 3$\sigma$ level (i.e. less
that 10$\%$).  Improved sampling is necessary to better constrain
variability.

We also have to note that, at higher frequencies, we lost flux density measurements of some components (see Tab. 3). This is a consequence of steep spectral index (see Tab. 4), low surface brightness of these (see Tab. 3) combined to probably not enough sensitivity of our images (see Tab. 2). Improved high quality images are necessary to give a complete spectral analysis of these components.

\subsection{The parsec-scale spectrum}

We used our multifrequency data from 2008 February 26 observations to
study the spectral index (defined $S_\nu \propto \nu^{-\alpha}$)
distribution of the parsec-scale structure.  For our analysis, we
first obtained images of the source using the same maximum and minimum
baseline in the ($u$, $v$) coverage, the same griding and the same
restoring beam. Then, we identified the different source components,
and we measured their flux densities.  Finally, we derived the
spectral index for each component between 1.6 GHz and 5 GHz, 5 GHz and
8.4 GHz and between 8.4 GHz and 22 GHz (see Table 4).  

\begin{table}[t!]
\caption{Spectral index for single component.}
\label{tab: index}
\centering
\begin{tabular}{cccccc}
\hline\hline
$\nu_1$-$\nu_2$ & HPBW& Component&S$_{\nu_1}$&S$_{\nu_2}$& $\alpha$\\
GHz &mas$\times$mas, $^{\circ}$ &&mJy&mJy&\\
\hline\hline
&&&&&\\
1.6-5&10.0$\times$5.0, 0&A&15.4&7.5& 0.63$\pm$0.01\\
&&B&29.0&3.6&1.82$\pm$0.01\\
&&N1&15.4&6.1&0.82$\pm$0.02\\
&&S1&9.4&7.3&0.66$\pm$0.03\\
&&&&&\\
5-8.4&3.0$\times$3.0,0&A&6.7&6.7&0.00$\pm$0.05\\
&&C&4.1&4.1&0.00$\pm$0.05\\
&&B&6.2&3.7&1.03$\pm$0.06\\
&&S1&7.7&2.6&2.12$\pm$0.06\\
&&&&&\\
8.4-22&1.86$\times$0.86,-4&C&5.3&4.6&0.14$\pm$0.04\\
&&A1&2.9&2.2&0.29$\pm$0.08\\
&&A2&8.1&6.8&0.18$\pm$0.03\\
&&&&&\\
\hline
&&&&&\\
\multicolumn{6}{l}{\scriptsize S$_{\nu_1}$ and S$_{\nu_2}$ are referred to the flux density at lower ${\nu_1}$ and higher ${\nu_2}$ frequency}\\
 \multicolumn{6}{l}{\scriptsize between the two considered for the spectral analysis. All flux density values are }\\
 \multicolumn{6}{l}{\scriptsize derived from 2008 February 26 observations. }
\end{tabular}
\end{table}

Between 1.6 GHz and 5 GHz owing to the low resolution, single
components are not well defined. The spectrum is in general steep as
expected from low brightness extended components. We note that the
steepest component is component B ($\alpha$ = 1.8), and that the
Northern extension (N1) is steeper, and more extended than S1.

At higher resolution, component C shows a flat spectrum
($\alpha_{5}^{8.4}$$\sim$0 , $\alpha_{8.4}^{22}$$\sim$0.14) and it
appears unresolved by a gaussian fit to the self-calibrated data. For
these reasons and because of the source morphology, we identify C as
the center of activity for 4C 26.42 with position RA 13$^{h}$ 48$^{m}$
52$^{s}$.4894321 $\pm$ 0.0000064,  DEC 26$^{d}$ 35$^{m}$ 34$^{s}$.340598
$\pm$ 0.000038 (estimated from the high resolution observations at 22 GHz
of 2008 February 26).  Also component A (with subcomponents A1 and A2)
shows a clear flattening in the spectrum at high frequency, while 
component B
despite the symmetric source morphology, has a steep spectrum
(1.0) between 5 and 8.4 GHz, and as expected is not detected at 22
GHz. We do not interpret this steep spectrum as due to different
synchrotron aging because the component N1 shows a steeper spectrum
with respect to S1.  Instead we suggest that the difference in 
spectral index is due to asymmetric interactions with the
surrounding medium in the bending region. We note that the source is
symmetric also on the kiloparsec-scale.

\begin{figure*}[t!]
\centering
\includegraphics[width=8.3 cm]{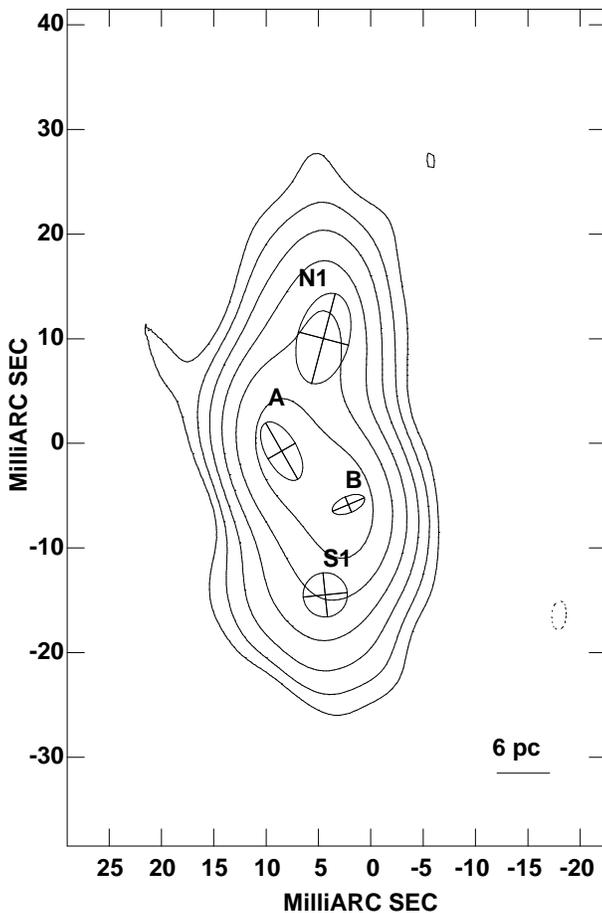}
\hfill
\includegraphics[width=7cm]{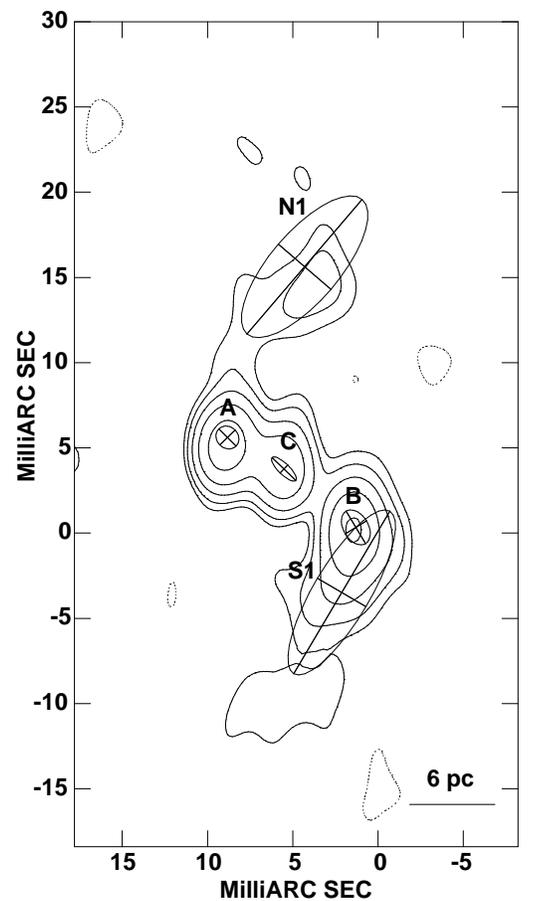}
\caption{On the left: 1.6 GHz VLBA contour map obtained on 2003 August 07 with the 
4 components given by modelfitting overlaid. 
The HPBW is 11.0 x 6.1 mas at 6.8$^{\circ}$ with a noise level = 0.15 mJy/beam . Contours are: -0.45, 0.45, 0.9, 1.8, 3.6, 7.2, 14.4 mJy/beam.  On the 
right: 5 GHz VLBA contour map obtained on 2005 July 27 with the 
5 components given by modelfitting overlaid. 
The HPBW is 3.4 $\times$ 2.2 mas at -8.0$^{\circ}$ with a noise level = 0.13 mJy/beam . Contours are: -0.4, 0.4, 0.8, 1.6, 3.2 and 6.4 mJy/beam.  }
\end{figure*}

\begin{figure}[t!]
\centering
\includegraphics[width=8cm]{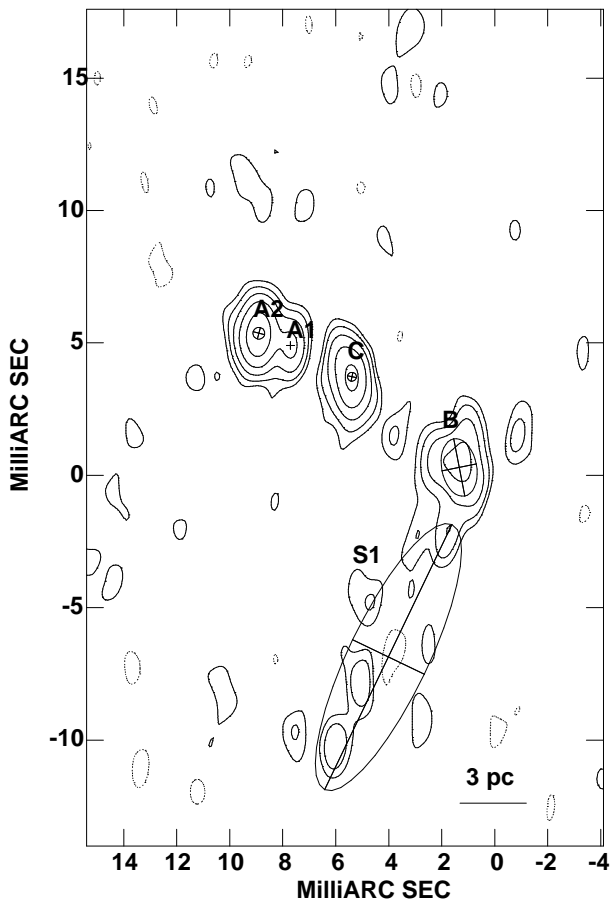}
\caption{ 8.4 GHz VLBA contour map at the 2008 February 26 epoch
with the 5 components derived from modelfitting overlaid. The HPBW is 1.86 $\times$ 0.86 
mas at -4.3$^{\circ}$ with a noise level = 0.08 mJy/beam. Contours are: -0.25, 0.25, 0.5, 1, 2, 4 mJy/beam.}
\label{fig:}
\end{figure}

\begin{figure}[h!]
\centering
\includegraphics[width=8cm]{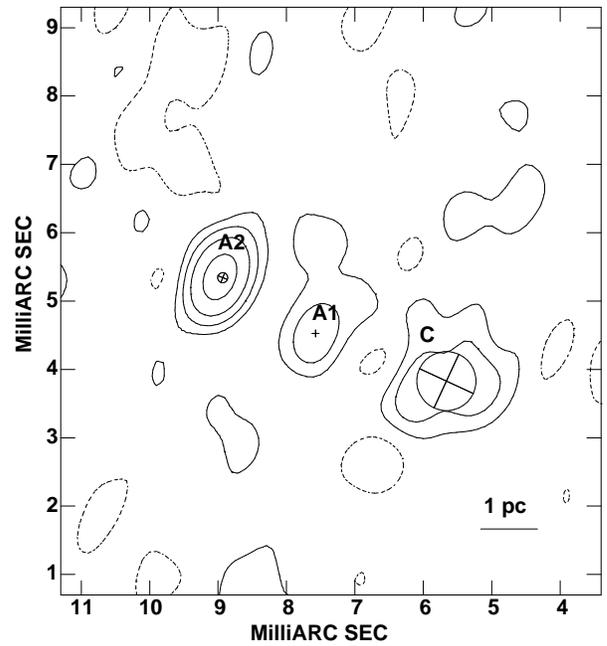}
\caption{22 GHz VLBA contour map obtained on 2008 February 26 with the 3 
components derived from modelfitting overlaid. The HPBW is 1.01 x 0.65 mas at 
-18.5$^{\circ}$ with a noise level = 0.15 mJy/beam. Contours are: -0.45, 0.45, 1.2, 2.4, 4.8 mJy/beam }
\label{fig:}
\end{figure}

\subsection{Jet dynamics}

\subsubsection{Proper Motion}

We studied the apparent proper motion of components A and B using the
three different epochs at 5 GHz. We have observations at two different
epochs also at 1.6 GHz, but the low resolution does not allow for a
reliable search for possible proper motion.  At 8.4 and 22 GHz only
one epoch is available. We note that components A and B are well
defined and their identification is not ambiguous, while components N1
and S1 are extended with no evident substructures so that their best fit
position is affected by the varying sensitivity of different
observations.

No significant proper motion was detected with a limits on the motion
of A and B with respect to C of $\sim$9$\times$10$^{-3}$ mas/yr
($\sim$3.6$\times$10$^{-2}$c) at the 3$\sigma$ level.  Available data
are poor (only 3 epochs) but with good time coverage (1997, 2005 and
2008 epochs). Therefore we conclude that A and B are stationary
features, probably not due to moving knots in the jets but instead
resulting from standing shocks in the region where the jet direction
undergoes a symmetric P.A. change of about 90$^\circ$.

\subsubsection{Bulk Velocity}

An estimate of the jet bulk velocity rather than pattern speed can be
derived from the jet to counter-jet brightness and size ratio, and by
comparison of the nuclear with the total radio power (see
e.g. \citet{gio88}).  The brightest component at 5 GHz at all epochs
is A which is the nearest to C. The flux density ratio with respect to
B at 5 GHz is S$_{A}$/S$_{B}$ $\sim$ 1.4. On the other hand, the
distance ratio is r$_{A}$/r$_{B}$ $\sim$ 0.8.  Moreover from an
inspection of available images we note that an underlying jet
connection between different structures (A, B and C) is not visible
probably because of sensitivity limitations. In conclusion, there is no
evidence of asymmetry in flux or position that would be expected
from Doppler boosting.

In most FR I or FR II sources with good VLBI data a relatively simple
nuclear-jet structure is present. The VLBI core shows a large
fraction of the arcsecond core flux density, and the correlation
between the core and total radio power can provide constraints on the
jet velocity and orientation.  In this source the arcsecond unresolved
core shows a complex and extended structure in VLBI images which are
not core dominated. The VLBI core is very faint and most of the flux
density is from jet-like components or small regions reminiscent of
radio lobes.  Therefore it is not possible to estimate the jet
velocity and orientation from the core dominance.

We can only note that the lack of underlying jets connecting the
different knots, the source symmetry and the low power of the nuclear
structure, all suggest that either the core structure of 4C26.42 is
near to the plane of the sky or/and the jets are not relativistic.
This conclusion reinforces the identification of mas structures as
standing shocks in the jet.
 
\subsection{The physical conditions at equipartition}

X-ray studies have shown that the black holes at the center of
galaxiy groups and clusters are capable of inflating cavities or
$``$bubbles$''$ in the
surrounding X-ray emitting gas \citep{fa03, fa05, bi04}.
The core of radio emission in 4C 26.42 is clearly identified with the
highest X-ray brightness region and therefore with the center of the
cluster.  Similar to many recent examples in literature, the
correspondence between the location of the radio lobes and X-ray
cavities \citep{john02} 
suggests an expansion of the radio lobes into the
surrounding X-ray emitting gas.  In this section we 
compute the magnetic field strength, internal pressure, and total energy, 
assuming that equipartition conditions apply.

We use the standard formulae in \citet{pac70}
, assuming that the 
relativistic particles and the magnetic field fully occupy the same volume 
($\Phi$ = 1), and that the amount of energy in heavy particles equals
that in electrons ($k$=1). We integrated over the frequency range 
10$^{7}$-10$^{11}$ Hz and we infer the extent of the source along 
the line-of-sight.

For the nuclear region imaged at 22 GHz (component C), 
we estimate a global 
equipartition magnetic field H$_{eq}\sim$0.22 Gauss. The total energy is 
E$_{tot}\sim$6.9$\times$10$^{52}$erg, 
while the corresponding minimum internal 
pressure is P$_{eq}\sim$2.4$\times$10$^{-3}$ dyn/cm$^{2}$.
For extended regions we used 1.6 GHz data and
for the lobe on the north  N1, we obtained 
H$_{eq}\sim$1.73$\times$10$^{-2}$ Gauss,  
E$_{tot}\sim$1.6$\times$10$^{53}$erg and 
P$_{eq}\sim$1.5$\times$10$^{-5}$ dyn/cm$^{2}$. 
Instead for the lobe on the south S1,  we 
obtained H$_{eq}\sim$4.28$\times$10$^{-2}$ Gauss, 
E$_{tot}\sim$2.9$\times$10$^{52}$erg and 
P$_{eq}\sim$8.9$\times$10$^{-5}$ dyn/cm$^{2}$.

\citet{bi04} estimated the energy required to inflate the large-scale 
bubbles in 4C 26.42. The energy required to create the observed 
bubble on the north in the X-ray emitting gas is E$_{bubble}$$\sim$39 $\times$10$^{57}$ 
ergs and the age of the bubble is $t_{age}$$\sim$1.8 $\times$ 10$^{7}$ yrs, 
where $t_{age}$ is R/$c_{s}$ where R is the distance of the bubble center 
from the black hole and $c_{s}$ is the adiabatic sound speed of the gas at 
the bubble radius. From these values we derive the required 
average jet mechanical power 
P$_{jet}$= E$_{bubble}$/$t_{age}$ involved in $``$blowing$''$ the bubble. We find 
P$_{jet}$$\sim$6.9$\times$10$^{43}$ergs/s ($\sim$ 6.9 $\times$10$^{36}$ W). 
Comparing the mechanical power of bubble with the heating necessary to 
prevent the gas from the cooling to low temperatures, \citet{bi04} found that the cavities can balance the cooling if they are relativistic and 
non-adiabatic, and
there may be further energy input if they are overpressured or
produce a shock when they are formed.
 
We can compare P$_{jet}$ derived above with the present parsec-scale
jet power.  Assuming for the northern jet a flux density upper limit
of 20 mJy at 5 GHz, its bolometric radio power is $\sim$8 $\times$
10$^{33}$ W, significantly lower than the power estimated from the
Northern cavity and therefore indicating radiatively inefficient jets
(0.1$\%$ of P$_{jet}$).  This result suggests that either the average
jet power is much higher, possibly because of the presence of heavy
particles, non-equipartition conditions, or an AGN activity that was
higher in the past.  We note that a heavy, slow jet is in agreement
with the results discussed in Section 3.4. We also point out that
low radiative efficiency seems to be a common theme for radio 
sources in clusters \citep{tay06b, al06}. 

\begin{figure*}
\centering
\includegraphics[width=16cm]{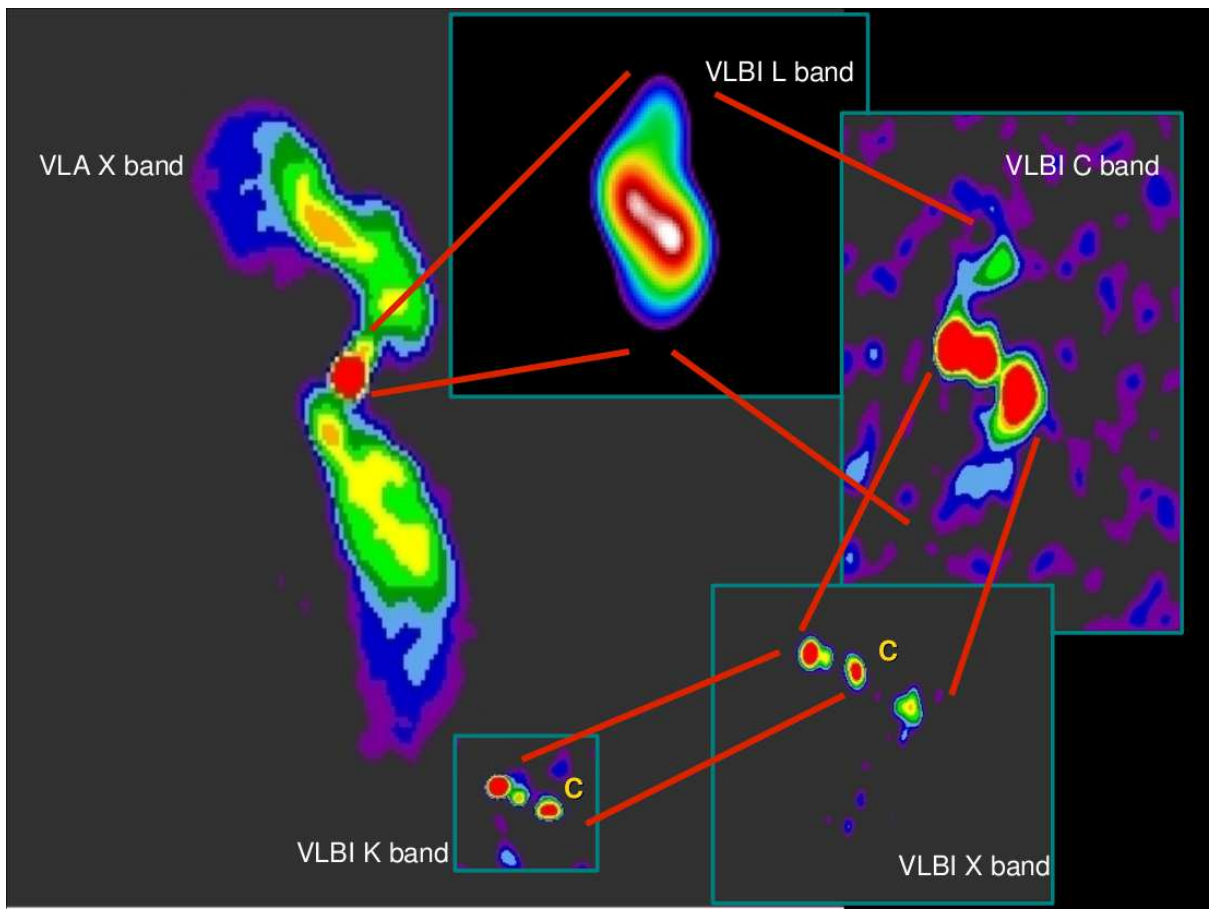}
\caption{Clockwise from left to right, zooming from kiloparsec to mas scale radiostructure of 4C 26.42: color maps of VLA X band, VLBI L band, VLBI C band, VLBI X band and VLBI K band data. (C) indicates the core component. }
\label{fig:}
\end{figure*}

\section{Discussion.}

The radio morphology on the parsec and kiloparsec scale of 4C 26.42 is
very complex. From spectral index considerations and modelfitting
results, we identify component C as the radio core. We note that the
jets bend by $\sim$90 $^{\circ}$ on scales of 10s of parsecs, and are
remarkably symmetric.  Two possibilities can explain the observed
morphology in 4C 26.42: 1) if the pc scale jets are relativistic as in
most radio galaxies (see e.g. \citet{gio01, gio05})
, and as suggested
by unified models, this source has to be oriented at $\sim$ 90$^\circ$
with respect to the line-of-sight. In this case the large symmetric
change in the source PA is real and not affected by projection. We
find it difficult for a highly relativistic jet to survive such a
large change in its orientation. Moreover we note that for a
relativistic jet in the plane of the sky the Doppler factor is much
lower than 1. Any decrease in the jet velocity will increase the
Doppler factor and thereby manifest as an enhancement in the surface
brightness. Since no clear surface brightness discontinuity is present
in our images, we conclude that the jet velocity before and after the
large PA change should be roughly the same.  2) Alternatively the
source morphology suggests a subsonic flow which might be expected to
have entrained thermal gas. This hypothesis is in agreement with the
result that data at three different epochs do not show any proper
motion, and with the identification of the bright regions on both
sides of the core as bright standing shocks.

The presence of non-relativistic jets in this source is in sharp
contrast with the observational evidence that parsec-scale jets in FR
I radio galaxies are relativistic \citep{gio01}.  
In FR I radio
galaxies relativistic jets in the parsec-scale region slow down
because of interaction with the ISM on the sub-kiloparsec
scale. However we must acknowledge the peculiar position and physical
conditions of 4C26.42 at the center of a cooling flow cluster.  In
Brightest Cluster Galaxies (BCGs) the presence of a dense ISM in the
central (parsec-scale) regions is expected since the cooling flow is
related to the presence of a high density gas as confirmed by the
detection of CO line emission and molecular hydrogen \citep{ed01,  ed02}. \citet{ro08} 
showed that a jet perturbation grows because of Kelvin-Helmotz instability and produces a
strong interaction of the jet with the external medium with a
consequent mixing and deceleration.  The deceleration is more
efficient increasing the density ratio between the ambient medium and
the jet. Relativistic light jets are expected in FR I sources and the
above effect can produce their slowing down from the parsec to the
kiloparsec scale as found in many sources \citep{tay96, ro08}. 
However a large value of the density ratio can produce a sub-relativistic and
heavy jet even on parsec-scales as suggested by our data on 4C 26.42.
 
A strong interaction with the ISM could also explain the large
difference in the spectral index distribution of the Northern and
Southern region.  The symmetric radio morphology suggests that the
different spectral index distribution is related to an anisotropy in
the ISM. We concluded in subsection 3.4.2 that the mas structures
result from standing shocks.  Internal shocks can arise from sudden
changes in the medium external to the jet.  These shocks cause
radiative losses through enhanced synchrotron radiation due to both increased particle energy density and magnetic field strength
behind the shock. In particular, we note from Table 4 that the southern part
(components B and S1) of 4C 26.42 is steeper than the northern part
(components A and N1). This behaviour could be due to a denser
ISM in the South than in the North region around our source.

We recall that in this source there are two dramatic changes in 
PA.  A similar, but reverse change is present at about 2$''$
from the nuclear region. The origin of this peculiar morphology is not
clear, but in any case requires the presence of heavy slow jets.  A
possible origin of a large symmetric change in the jet direction could
be the interaction with a rotating disk as suggested by \citet{van84}
for 3C 293. In this model, a slow jet could be bent by ram pressure of
a rotational motion of the accreted gas.  The ISM of the galaxy
rotates supersonically, so there is a stand-off cylindrical bow shock
in the interstellar gas upstream of the jet.  A result consistent with
this scenario is the spettroscopic discovery in the nuclear region of 4C 26.42 of two clouds with a
difference velocity of $\sim$ 330 km/sec. Their emission line ratios of H$\alpha$ and H$\beta$ are very similar and typical of Liners exceptionally weak O [III] lines \citep{gon99}. 
However, no clear evidence of rotating gas has been found. Moreover, the second
change in direction would require two counter-rotating regions one
near to the core and one at larger distance to explain the two changes
in the jet direction.

An alternative possibility is that on the large scale the radio lobes
rise buoyantly in the N-S direction according to the thermal gas
distribution (see e.g. the radio X-ray overlay presented by
\citet{fa01}.  On intermediate scales, the jets have been heading out
at P.A.  $-$30$^{\circ}$ for a long time. Then, on the smallest scale,
we can suppose that the radio jets have just been realigned recently
due to the merger of a Binary Black Hole system which could produce a
change in the nuclear jet ejection direction. \citet{mer02} suggest that the orientation of
a black hole's spin axis could change dramatically due to a
merger event, leading to a sudden flip in the jet direction.  Merger
events are relatively common in cD galaxies, and in fact \citet{john91} proposed that the structure in the envelope of the
cD galaxy in A1795 is due to a merger with another giant galaxy.  Alternatively the
change of the jet PA on the parsec-scale could be due to a complex angular momentum of the gas accreted in the inner part of
the disk of this galaxy, as proposed by \citet{au06} 
for B2151+174
(BCG of Abell 2390) or by rotating vortices in the cluster gas as \citet{fo09} 
suggested for the unique morphology of 3C 28, the BCG of
Abell 115.

\section{Conclusion.}

New sensitive, high resolution images at 1.6, 5, 8.4 and 22 GHz are
presented for 4C 26.42, the radio loud BCG at the center of the
cooling flow cluster A1795. Our multi-frequency and multi-epoch VLBA
observations reveal a complex, reflection-symmetric morphology over a
scale of a few mas. No significant proper motion of observed components
is found (\S3.4). We identify the core (component C), along with two
stationary shocks (A and B) and two lobes (N1 and S1). A strong
interaction with the ISM can explain the spectral index distribution
witch is steeper in the southern region of the source,  and the 
presence of sub-relativistic and heavy jets on the parsec scale (\S 4).  
We note two 90$^\circ$ jet changes: 
one at $\sim$ 4-5 mas and one at 2$''$ from the nucleus. 
It seems plausible that the two
90$^{\circ}$ P.A. changes may have a different origin: we suggest
buoyancy effects on the large scale, while on the mas scale it is
difficult to identify if SMBH precession or peculiar gas motions are
able to produce the sudden change in direction. To
better investigate the peculiar structure of 4C 26.42, it would be
useful to look at this source with a resolution in between that of the
VLBA and VLA, as for example combining EVLA and e-Merlin data.

\begin{acknowledgements}
We thank the staff of NRAO involved in the observations for their
help. NRAO is a facility of the National Science Foundation, operated
under cooperative agreement by Associated Universities, Inc. This
research has made use of the NASA/IPAC Extragalactic Data Base (NED),
which is operated by the JPL, California Institute of Technology,
under contract with the National Aeronautics and Space Administration. This research was partially supported by ASI-INAF CONTRACT I/088/6/0.
\end{acknowledgements}

\end{document}